\documentclass{vldb}
\usepackage{graphicx}
\usepackage{balance}  

\newcommand{\comment}[1]{\mbox{}}

\begin{document}
\title{Detecting dense communities in large social and information networks with the Core \& Peel algorithm}

%
\numberofauthors{3}
\author{
%
%
\alignauthor
Marco Pellegrini\\
       \affaddr{Institute for Informatics and Telematics}\\
       \affaddr{Consiglio Nazionale delle Ricerche}\\
       \affaddr{Via Moruzzi 1, 56124 Pisa, Italy}\\
       \email{m.pellegrini@iit.cnr.it}
\alignauthor Filippo Geraci \\
\affaddr{Institute for Informatics and Telematics}\\
       \affaddr{Consiglio Nazionale delle Ricerche}\\
       \affaddr{Via Moruzzi 1, 56124 Pisa, Italy}\\
       \email{f.geraci@iit.cnr.it}
\alignauthor
 Miriam Baglioni \\
\affaddr{Institute for Informatics and Telematics}\\
       \affaddr{Consiglio Nazionale delle Ricerche}\\
       \affaddr{Via Moruzzi 1, 56124 Pisa, Italy}\\
       \email{m.baglioni@iit.cnr.it}
}
\comment{
\author{\IEEEauthorblockN{Miriam Baglioni\IEEEauthorrefmark{1},
Filippo Geraci\IEEEauthorrefmark{1},
Marco Pellegrini\IEEEauthorrefmark{1}and
Ernesto Lastres\IEEEauthorrefmark{2}
}
\IEEEauthorblockA{\IEEEauthorrefmark{1}IIT - CNR, Pisa, Italy}
\IEEEauthorblockA{\IEEEauthorrefmark{2}Sistemi Territoriali, Navacchio (Pisa), Italy}}
}
\maketitle

\begin{abstract}
Detecting  and characterizing   dense subgraphs
(tight communities) in social and information networks  is an
important exploratory tool in social network analysis. Several
approaches have been proposed that either (i) partition the whole
network into "clusters", even in low density region, or (ii) are
aimed at finding a single densest community (and need to be iterated
to find the next one). As social networks grow larger  both
approaches (i) and (ii) result in algorithms too slow to be
practical, in particular when speed in analyzing the data  is
required. In this paper we propose an approach that aims at
balancing efficiency of computation and expressiveness/manageability
of the output community representation. We define the notion of a
partial dense cover  (PDC) of a graph. Intuitively a PDC of a graph
is a collection of sets of nodes that (a) each set forms a disjoint
dense induced subgraphs and (b) its removal leaves the residual
graph without dense regions. Exact computation of PDC is an
NP-complete problem, thus, we propose an efficient heuristic
algorithms for computing a PDC which we christen {\em Core \& Peel}.
Moreover we propose a novel benchmarking technique that allows us to
evaluate algorithms for computing PDC using the classical IR
concepts of precision and recall even without a golden standard.
Tests on 25 social and technological networks from the Stanford
Large Network Dataset Collection confirm that Core \& Peel is
efficient and attains very high precison and recall.
\end{abstract}

\section{Introduction}
\label{sec:intro}

Finding dense sub-networks in complex (social, communication,
technological, biological) networks is a key exploratory tool in
complex network analysis as often one can spot emerging phenomena
liked to these regions with an above-average level of activity. As
we will use social networks as a key metaphor we will talk of dense
{\em communities} of nodes.

{\bf A brief taxonomy of approaches to community detection}. The
recent literature the detection of community structures in complex
graphs can be roughly split into three trends.

The fist approach has concentrated on {\em partitioning} the nodes
of a graph into clusters (or modules) so to favor the formation of
modules for which intra-modules connections are favored over
inter-module connections (e.g. the classical algorithm of Girvan and
Newman  \cite{GiNe02}). This approach uses a non-local definition of
"density" and postulates that every node should belong to some
community (or alternatively) that all communities even if a low
density or hard to distinguish from the background are of interest
(approach A). Being a partition of the nodes, the communities found
are disjoint.

The second approach  postulates that  a connected community
maximizing the density should be sought
\cite{Andersen2009,Charikar2000}. A collection of dense communities
can be constructed by removing the community found from the input
graph and iterating the method on the residual graph (approach B).
Communities found in this way are disjoint.

A third approach aims at enumerating all dense subgraphs within the
graph, usually with a condition of maximality, but allowing overlaps
among communities  (see e.g. PalEtAl05)(approach C). In the worst
case, the size of the output of an algorithm listing all maximal
cliques (or maximal dense subgraphs) can be exponentially larger
than the size of the input.

Any of the aforementioned approaches has its pros and cons. Very
Large social networks currently available pose a challenge to
community analysis approaches in terms of computation time, main
memory storage and readability of the algorithms' results.

So while approaches (A) and (C) may reveal a complex  community
landscape, they do so in a quite unfocussed manner, that implies
slow computations, and/or high memory consumption, and  an output
that may need further analysis/modelling/visualization to be of use.
In contrast approach (B) may be fast if appropriate heuristics are
employed, but reveals only a single community in an otherwise more
complex landscape. A the simple solution consisting in   iterating
approach (B) over the residual graph obtained by removing the found
densest community would be an overshooting for this problem, taking
us back to the problem of an excessive time complexity.

The fourth approach, which is advocated in this paper as well as in
others (see e.g. \cite{KumarRRT99}, \cite{DourisboureGP09},
\cite{Gibson05}
 \cite{Chen10}, \cite{Wang2008}) is that of collecting in one go a collection of
disjoint dense communities above a certain size and density
thresholds. This approach is suitable for very large graphs since
the output is easy to interpret, visualize, and feed to further
filters and analytic tools, in particular it is often the method of
choice for the web graphs because of their size.

{\bf Our contribution.} This paper presents two main contributions.
The first is a formalization of the problem of (disjoint)
communities detection as that of computing a {\em partial dense
cover of a graph} and an efficient algorithm (christened {\em Core
\& Peel}) for computing heuristically the  partial dense covers of a
graph. The second contribution is methodological and aims at
improving our capability of measuring the
 quality of an heuristic algorithms for (disjoint) community detection.
In particular we propose a scheme that allow us to measure
performance by classical IR concepts of {\em precision} and {\em
recall} or realistic (i.e. non-random) graphs even without using a
golden standard.

{\bf The Core \& Peel algorithm}. We present an heuristic algorithm
for detecting all sufficiently dense subgraphs of a graph $G$ whose
removal produces a residual graph $G'$ without dense subgraphs. The
main idea consists in visiting  the nodes of a graph in an
appropriate order and for each visited node (called a seed) we look
in the neighborhood of the seed in order to isolate  those nodes in
the neighborhood that, together with the seed, can participate in an
a sufficiently dense  induced subgraph. The order of the visit of
the node is chosen so to visit first those nodes more likely to
contribute in a larger dense subgraph. Large and  dense subgraph
must be detected only once so to avoid wasting time re-discovering
the same subgraph multiple times (i.e. with  by a single seed) thus
once such dense subgraph is detected its component vertices are
marked and cannot be used as seeds any more. The visit order is
produced by computing first the {\em core number} of each node in
the graph, which represent an upper bound to the average degree of
the largest clique (or quasi-clique) incident to a node. Once a
suitable region around a seed is detected that is likely to hold a
dense subgraph, we employ a variant of the {\em peeling} procedure
described by Charikar in \cite{Charikar2000} to extract the dense
subgraph. The key definitions and the algorithm is described in
\ref{sec:heuristic}.

{\bf Benchmarking methodology}. In may areas of computer science the
lack of a common benchmarking methodology is often recognized as one
of the main obstacles towards speedy  progress in solving the key
problems of that area. The IR community has been one of the first
success stories in setting up an effective benchmarking methodology
codified through the TREC conference series. For the problem of
clique detection the TCS and OR communities established the DIMACS
benchmark collection. While many methods for community detection
have been proposed recently, there is still no accepted benchmarking
methodology. Most papers address the issue of algorithmic validation
by planting a few dense subgraph in a random graph and comparing the
predicted dense subgraph against the planted ones. This "standard"
approach is however wanting in many respects.

(a) First of all it is well known that the basic Erd\"{o}s-R\'{e}nyi
random graph model is not suitable for modelling complex (social,
technological, etc..) networks, while on the other hand it is still
an open area of research that of producing suitable generative graph
models. At the moment no single model can cover the whole range of
applications.

(b) Secondly, the embedded communities should not skew too much the
underlying graph distribution (e.g. the vertex degree distribution)
so to avoid making the instances amenable to approach based on
simple statistical filters.

(c) Thirdly, the possible presence of native (non-planted)  dense
subgraphs complicates the assessment as any method since detecting
such nodes cannot be considered a fault of the algorithm, and should
be weighted positively rather than ignored.

In section \ref{sec:testing} we propose a method that copes with
issues (a),(b) and (c) and results in a more stringent and realistic
assessment for community detection algorithms. In particular we are
able to define both precision and recall measurements that are
standard and well understood measures of quality in IR applications.

{

\section{Previous work}
\label{sec:previouswork}

Most of the known results are for finding a single maximum size
dense subgraph (or clique).  In \cite{lee2010} V.E. Lee et al. give
a survey of the state of the art on algorithms for dense subgraph
discovery. Here we recall the main known results.

{\bf Densest subgraph detection.} Dense subgraphs can be defined in
a variety of ways: if we seek the subgraph $H$  of $G$ maximizing
the average degree $2|E(H)|/|V(H)|$, then there is an exact
polynomial algorithm based on max-flow/min-cut computations
\cite{Gallo89,Goldberg84}. A second exact polynomial time algorithm
based on a reduction to LP is shown in \cite{Charikar2000}. A simple
greedy (1/2)-approximation algorithm\footnote{A $\rho$-approximation
algorithm $A$ for a {\em maximization problem}, with $ 0 \leq \rho
\leq 1$ is an algorithm that, calling OPT the value of the objective
function $f(.) \geq 0$ for the optimal solution, returns a solution
$x$, whose value of the objective function $f(x)$ is such that $\rho
OPT \leq f(x) \leq OPT$.} is also shown in \cite{Charikar2000}.

If we look for the densest subgraph with exactly k-vertices ({\tt
dks}), which is an NP-complete problem, Feige et al. \cite{Feige01}
gave an $O(n^{-1/3 +\epsilon})$-approximation algorithm, where
$n=|V|$, and $\epsilon>0$ an arbitrarily small positive constant.
The approximation coefficient has been later improved in
\cite{Bhaskara10} to $O(n^{-1/4 +\epsilon})$. It is known that this
problem does not admit a PTAS, unless some hard complexity
hypothesis  holds \cite{Khot06}.

If we look for the densest subgraph with {\em at least} k-vertices
({\tt dalks}) Andersen et al. \cite{Andersen2009} give an efficient
$(1/3)$-approximation algorithm for this problem, based on core
decomposition. In the same paper it is shown that the densest {\em
at most} k problem ({\tt damks}) is as hard as {\tt dks}. Somewhat
more complex (still polynomial) algorithms (based on LP and/or
max-flow) attain an approximation factor of $1/2$ for {\tt dalks}
\cite{Andersen07,KhullerS09}.

A different notion of "density" is used in the definition of
$\gamma$-quasi cliques. A graph $G=(V,E)$ is a $\gamma$-quasi clique
if  every node in $G$ is adjacent to at least $\gamma(|V|-1)$ nodes
in $G$.  The densest subgraph problem (with this notion of density)
can be recast as the problem of finding the largest induced subgraph
that is a $\gamma$-quasi clique, for a fixed value of $\gamma$.
Heuristics for detecting quasi-cliques are developed in
\cite{Liu08,Abello02}. A method for detecting quasi-cliques tailored
for external memory computations is given in \cite{ZengWZK07}.
Since, for $\gamma = 1$ we obtain a clique, this problem is in
general as hard as maximum clique detection.

When the subgraph  sought are relatively large and dense, a sampling
approach based on shingling by Gibson et al. \cite{Gibson05} works
well in practice.

In \cite{WangZTT10} Wang et al. define dense subgraph by placing a
lower bound on the number of common neighbors for each pair of
adjacent nodes in the subgraph, and propose a method  based on
iterating an efficient triangle counting black-box algorithm.

\comment{ Mishra et al. \cite{MishraSST08} propose to use the notion
of $(\alpha,\beta)$-cluster that simultaneously gives a lower bound
$\alpha$ to the average degree in the induced subgraph, and a lower
bound $\beta$ to the average number of cut edges joining the
subgraph with its complement in the input graph.}

{\bf Maximum clique.}  The maximum clique problem  for a graph G
 requires to find the the largest
complete subgraph of $G$, and is a classical NP-complete problem.
Since this problem is also hard to approximate \cite{Hastad96} the
research effort has been directed towards heuristic, or exact
algorithms still exponential in the worst case, but able to be fast
in practice over certain classes of input graphs (see \cite{Bomze99}
for a survey).

{\bf Other graphs.} In this paper we restrict ourselves to  generic
(undirected) graphs. There is also a  vast literature  which we do
not touch here (and corresponding different definitions) for
directed graphs, for bipartite graphs, and other generalizations or
specializations of graphs.

{\bf Dense subgraphs and communities in social networks}.  In social
graphs often communities are defined in terms of dense subgraphs
(see \cite{fortunato2010,GiNe02,newman03structure,Kosub04}. Also
dense bipartite subgraphs are important in some applications since
web communities are often modeled a as bipartite subgraphs of hubs
and authorities (see e.g. \cite{KumarRRT99,DourisboureGP09}).

{\bf Partial dense covers approaches.}  A work that has similar aims
as ours  is in \cite{Chen10}, where a partial vertex cover with
dense subgraphs is sought. In \cite{Wang2008} wang et al. map the
problem into a real geometric space and look for dense clusters of
points in this space.

{\bf Core Decomposition}. Since we use heavily properties of the
{\em core decomposition} of a graph, we survey some related results.
The k-core of a graph $G=(V,E)$ is defined in \cite{Seidman83} as an
 induced  subgraph $H$ of $G$, of minimum degree at least
$k$. The core number of a vertex $v$ of $G$ is the order of the
largest $k$-core to which $v$ belongs. The core decomposition of a
graph is a partition of the nodes into equivalence classes by the
core number of its vertices. The core number $cn(G)$ of a a graph is
the maximum core number of any vertex in $v$.

A very efficient method (linear in the graph size $O(|V|+|E|)$)  for
computing the core decomposition of a graph was proposed in
\cite{batagelj2003}. For a graph too large to fit in RAM memory a
method in \cite{Cheng11} allows to compute the core decomposition of
a graph with $O(cn(G))$ scans of the data. Several applications of
core decomposition to the analysis of networks are listed in
\cite{Cheng11}.

The core decomposition is used as a preliminary phase in the
approximate algorithms in \cite{Andersen2009}  for the densest
at-least-k subgraph problem ({\tt dalk}).

\section{Preliminaries}
\label{sec:preliminaries}

\noindent Let $G=(V,E \subset V \times V)$ be a simple (undirected)
graph (no self-loops, no multiple edges).

\noindent A subset $Q \subset V$ induces a subgraph $H_{Q} = (Q,
E_Q)$, where $E_Q  = \{ (a,b) \in E | a \in Q \wedge b \in Q \}$.

A nice survey of concepts and algorithms related to local density of
subgraphs is in \cite{Kosub04}. We restrict ourselves to local
density definitions, that are those for which the density of a
subset $Q$ is a function depending on  $Q$ only and $E_Q$.

\noindent For a graph $G$ is {\em average degree} is:

\[ av(G) = \frac{2|E|}{|V|}, \]

\noindent thus the ratio $|E|/|V|$ is just half the average degree.

\noindent Define as {\em density} of a graph $D(G)$ the following
ratio:

\[ D(G) = \frac{|E|}{\binom{|V|}{2}} = \frac{2|E|}{|V|(|V|-1)}, \]

\noindent which gives the ratio of the number of edges in $G$ to the
maximum possible number of edges in a complete graph with the same
number of nodes.

\noindent Cliques are subgraphs of density 1, and finding a maximum
induced clique in a graph $G$ is an NP-complete problem. Several
relaxations of the notion of clique have been proposed (see
\cite{Bala2007}) for a survey), most of which also lead to
NP-complete decision problems.

\noindent In the data mining literature it ha emerged the  concept
of quasi-clique. Given a parameter $\gamma \in [0..1]$, a
$\gamma$-quasi clique is a graph $G=(V,E)$ such that:

\[ \forall v \in V \ \ |N_G(v)| \geq \gamma(|V|-1), \]

where $N(G(v)= \{ u \in V | (v,u) \in E \}$ is the set of neighbors
of $v$ in $G$.

\noindent Note that a $\gamma$-quasi clique has density $D(G) \geq
\gamma$. In general however for a dense graph with density $D(G)$ we
cannot infer a bound on the value  of $\gamma$ for which it is a
quasi-clique (except for the value $D(G) =1$ that implies $\gamma
=1$, and those cases covered by Tur'{a}n's and Dirac's theorems
(\cite{Turan1941, Dirac1963}.

\noindent If we impose a lower bound to the  number of nodes  $k$ of
vertices in a subgraph, then the average degree and the density
depend only on the number of edges, and the attain maximum value for
the same graph. Otherwise finding the subgraph of maximum average
degree, or the subgraph of maximum density are quite different
problems, the latter admitting a polynomial time solution, the
latter NP-complete.

\noindent In this paper we aim at detecting dense-subgraphs with a
lower bounds on the size of the graph and to its density. However
the heuristic we propose is best understood by referring to the case
of cliques and quasi-cliques. Isolated cliques and quasi-cliques
have a  regular structure for which it is easy to derive useful
properties. When we look ad the induced cliques and induced
quasi-cliques in a graph, we cannot guarantee that those properties
are preserved, although they are in general rather robust with
respect to the perturbations produced by the other nodes, and thus
are often preserved in real-life social graphs.

\section{Heuristic for dense cover}
\label{sec:heuristic}
\label{sec:heuristic}

\subsection{Partial dense cover of a graph}

In this paper we propose an alternative approach that aims at
balancing efficiency of computation and expressiveness/manageability
of the output community representation. Our aim is to compute for a
graph $G = (V,E)$ and parameters $q \in [1,..|V|]$ and density
parameter $\delta \in [0.0,..1.0]$, a {\em partial dense cover}
$PDC(G)$ of $G$. We denote by $D[G]$ the density of $G$, i.e the
ratio of $|E|$ and $|V|(|V|-1)/2$.

A partial dense cover $PDG(G)$ is a  collection of subsets of $V$,
$\{C_1,...,C_k\}$. With the following properties

\begin{itemize} \item[a] $\forall j \in [1..k], \ C_j \subseteq V$
\item[b] {\em Node-disjointness:} $\forall j,h \in [1..k], \  j \neq
h  \Rightarrow C_j \cap C_k  = \emptyset$. \item[c] {\em Large size
and density}. $\forall j \in [1..k], \ |C_j| \geq q$, and $\forall j
\in [1..k], \ D[G[C_j]] \geq \delta$. \item[d] {\em Internal
maximality:} The union of any two sets $C_i$, and $C_j$ induces a
subgraph satisfying the other conditions and of density lower than
the minimum density of $G[C_i]$ and $G[C_j]$. \item[e] {\em External
maximality:} The residual graph $G[V\setminus \bigcup_{i=1}^k C_i]$
does not contain any induced subgraph satisfying the other
conditions. \item[f] {\em Radius constraint:} Each subgraph $G[C_i]$
has radius 1 or 2. \end{itemize}

The best way to justify the notion of a partial dense cover is to
consider it as a relaxation of the well known concept of a {\em
minimum clique partition} (problem GT15 in \cite{GareyJ79}. A
minimum clique partition  of a graph $G=(V,E)$ is a partition of the
vertex set $V$ into $k$  disjoint subsets so that the graph induced
in $G$ by each subset is a clique. Computing a clique cover of
minimum size ($k$) is a well known NP-complete problem
\cite{GareyJ79}, and it is hard to approximate \cite{Lund94}.

In a  clique cover every clique satisfies the  radius constraint
(f), as any clique has radius 1, however  graphs of radius 1 are not
cliques in general. The radius constraint to 1 or 2 is set in order
to attain an efficient computation, but as we will see later, we
will be able to cover a wide range of non-trivial communities.

The cover we seek is {\em partial} since we do not insist that any
vertex must belong to some covering set, however we ask in (e) that
the residual graph must not contain a subgraph that may be part of
the cover. Note that this condition (d) implies that the problem we
define is still NP-hard. In particular, for the density parameter
$\delta = 1.0$, our problem reduces to assessing whether a graph
contains no clique of size al least $q$, which is equivalent to the
NP-complete maximum clique problem.

The internal maximality constraint (d) is not problematic from a
computational point of view since it can always be enforces "a
posteriori" in polynomial time starting from a cover that satisfies
the other constraints.

For $\delta = 1.0$ and $q=1$ the problem is  reduced to that of
computing a {\em minimal clique cover}, that is, a clique cover
which cannot be improved by  locally merge two sets in the cover.
This extreme case can be solved polynomially, but, naturally  it is
different from the minimum cover problem.

The size parameter $q$ and density parameter $\delta$ ensure that we
can focus the computational effort towards those part of the graph
that are more interesting (i.e. of large size and high density) with
the aim of attaining computational efficiency and controlling the
amount of information gathered in a run of the algorithm.

\subsection{An Algorithm to compute partial dense covers}

As noted above computing a partial dense cover PDC(G) of a graph is
an NP complete problem. In this section we will give an efficient
heuristic algorithm which is based on combining in a novel way
several algorithms  and procedure already present separately in
literature. For each step we will give "hand weaving" arguments as
to it role and an intuitive reason as to why it contributes to
solving the PDC(G) computation efficiently.

\subsection{Heuristic Algorithm Core \& Peel}

{\em Phase I}. Initially we compute the Core Decomposition of $G$ (
denoted with $CD(G)$). Moreover  we compute for each vertex $v$ in
$G$ the {\em Core Count} of $v$ , denoted with $CC(v)$ , defined as
the number of neighbors of $v$ having core number at least as large
as $C(v)$. Next, we sort the vertices of $V$ in decreasing
lexicographic order of their core values $C(v)$ and core count value
$CC(v)$.


{\em Phase II}. In phase II we  consider each node $v$ in turn, in
the order given by phase I. For each $v$ we construct the set
$N_{C(v)}^r(v)$ of neighbors of $v$ at distance $r$ in $G$ having
core number at least $C(v)$. If $|N_{C(v)}^r(v)| < q$ we do not
process this node any more. Otherwise we compute the density
$\delta(v)$ of the induced subgraph $G[N_{C(v)}^r(v)]$. If this
density  is too small: $\delta(v) \leq \delta_{low}$, for a user
defined threshold $\delta_{low}$  we do not process this node any
more. Nodes that pass the size and density tests go to phase III.

{\em Phase III}. In this phase we take $v$ and the induced subgraph
$G[N_{C(v)}^r(v)]$ and we apply a variant of the  peeling procedure
described in \cite{Charikar2000}, that iteratively removes modes if
minim degree in the graph. The peeling procedure stops (and report
failure) when the number of nodes drops below the threshold $q$. The
peeling procedure stops (and reports success) when the density of
the resulting subgraph is above or equal to the user defined
threshold $\delta$. The set of nodes returned by the successful
peeling procedure is added to the output cover set, and its elements
are marked so to be excluded from further consideration.

Phases II and III are executed interleaved so to (1) enforce that
the covering sets produced are disjoint, and (2)  avoid discovering
multiple times the same dense subgraph.

\subsection{Justifications of phase I}

The core decomposition of a graph $G=(V,E)$ associates to any vertex
$v$ a number $C(v)$ which is the largest number such that $v$ has al
least $C(v)$ neighbors having core number at least $C(v)$.

Consider now a clique $K_x$ if size $x$, for each node $v$ in $K_x$
its core number is $x-1$. If $K_x$ is an induced subgraph of $G$,
then its core number can only be larger than $x-1$, thus $C(v)$ is
an upper bound to the size of the largest induced clique incident to
$v$.

Consider a quasi-clique $K_{x,\gamma}$, for each node $v$ in $K_x$
its core number is at least $\gamma (x-1)$. If $K_{x, \gamma}$ is an
induced subgraph of $G$, then its core number can only be larger,
thus  thus $C(v)$ is an upper bound to the size of the largest (in
terms of average degree) quasi-clique incident to $v$.

Thus if the upper bound provided by the core number is tight,
examining the nodes in (decreasing) order of their core number
allows us to home in first on the largest cliques (or
quasi-cliques), and subsequently the smaller ones.

In a clique $K_x$ each node is a {\em leader} for the clique,
meaning that it is at distance 1 to any other node in the clique.
Thus the first node of $K_x$ encountered in the order computed in
phase I is always a leader. In the case of of quasi-cliques we just
suppose the existence of at least one leader node. For an isolated
quasi-clique the leader node will have the the maximum possible core
count value, thus by sorting (in the lexicographic order) on the
core count value we push the leader node to be discovered first in
the order (assuming all nodes in the quasi-clique have the same core
number). For an induced quasi-clique the influence of other nodes
may increase the value of the core count for any node, but, assuming
that the relative order between the leader and the other nodes does
not change, we still obtain the effect of encountering the leader
before the other nodes of the quasi-clique.

The core number gives us an estimate of the largest (in terms of
average degree) quasi-clique (or clique) a node can be incident to,
thus it provides a very powerful filter. Also fortunately there is a
very efficient algorithm of complexity $O(|V|+|E|)$ that computes
the core decomposition of a graph \cite{batagelj2003}.

Also it has been observed empirically that in social graph often the
core number and the core count of any node are much smaller than its
degree (which is also an obvious upper bound on the size of the
largest clique incident to $v$). Thus subsequent operations are
intuitively  less expensive when their complexity can be charged
onto the core number or the core count values of a node and its
neighbors.

\subsection{Justification of phase II}

In Phase II we count the number of edges in the subgraph induced by
the node $v$ and its neighbours with equal or higher core value.
Phase II is equivalent to computing a {\em restricted local
clustering coefficient} of a node $v$,  where the computation is
restricted to the neighboring nodes of $v$ of sufficiently high core
number.

Classical result of Turan and Dirac (see \cite{Turan1941},
\cite{Dirac1963}) guarantee the existence of a clique (or a clique
with a few edges missing) in graphs with sufficiently many edges.
(approximately above $n^2/4$ for a graph of $n$ nodes). Although it
is possible to derive some  thresholds from   these classical
results, such threshold would give us only sufficient conditions but
not necessary ones to the existence of a clique in the graph that is
exhamined in phase II. Since the purpose of phase II is to trade off
the number of invocations of phase III with the chance that phase
III may find a sufficiently dense subgraph, we resort to a more
ad-hoc strategy for selecting thresholds. We set $\delta_{low} =
\delta/2$, which did perform well in our experiments.

\subsection{Justification of phase III}

The peeling procedure we use is similar to one described in
\cite{Charikar2000}. It consists in an iterative procedure that
remove a node of minimum degree (ties resolved arbitrarily) and all
incident edges, then iterates on the residual graph. In
\cite{Charikar2000} the  graph of highest average degree constructed
in this process is returned as output. We modify this procedure by
returning the first subgraph generated that satisfies the density
and size constraints.

It is shown in \cite{Charikar2000} that this procedure is
(1/2)-approximate for the {\em average degree} , i.e. it returns a
subgraph whose average degree  is with a factor (1/2) of that of the
subgraph of highest average degree. Empirically, we rely on the
intuition that, in our setting, the input to the peeling procedure
as produced after phase II is a superset of the output that is tight
enough and dense enough so that the peeling procedure converges
quickly to isolating the embedded dense subgraph, in a typical
situation.

We also use a novel heuristic to solve cases of ties within the
peeling algorithm in \cite{Charikar2000}. When two or more vertices
are of minimum degree the original peeling procedure picks one
arbitrarily. In our variant we compute the sum of degrees of the
adjacent nodes $D(v) = \sum_{w \in N(v)} d(w)$ and we select the
vertex among those of minimum degree minimizing $D(.)$. This
secondary selection  criterion is inspired by observations in
\cite{HalldorssonR94}, where the objective is to select an
independent set by iteratively removing small degree nodes, which is
a dual problem that of detecting cliques.

\subsection{Influence of radii}

In our algorithm we assume that the dense community has radius $r=1$
or $r=2$. This technical restriction is needed in order to ensure
that  we can recover efficiently a superset of the community node
set starting from a leader seed node. As we will argue below this
restriction is not severe as it covers a wide range of cases of
interest either precisely or with high probability.

{\bf Case $r=1$}. As mentioned above  in a clique $K_n$ of $n$ nodes
every node is at distance 1 from any other node, thus every node is
a center and the radius is $r=1$. If we start removing edges from
$K_n$ we maintain the property $r=1$ with certainty up to $\lfloor
n/2 \rfloor -1$ edge removals, while the number of centers is
reduced. If we look at the edge removal as a random uniform
selection of a pair of nodes and we remove of the edge joining them
(if still existing) we can model  this process similarly to the well
known {\em Coupon collector problem} \cite{MotwaniR95} thus the
expected number of edge deletions before all nodes are selected at
least once is approximately $(n/2)\ln n$. Thus even for a fair
relaxation from the clique we retain the property with high
probability of to this limit since
 the Coupon Collector number has a distribution highly concentrated around the expected value.

In certain applications a node that represents a center (realizes
$r=1$) in a dense subgraph can be interpreted as a "leader" of the
group (since it has relationships with all the members of the group)
and thus being a feature characterizing the proper communities. In
effect in many social graphs (e.g. Twitter graph) there is a natural
leader-follower dynamics.

{\bf Case $r=2$}. If we remove edges but the degree of each node is
at least $\lfloor n/2 \rfloor$ then by the pigeon hole principle
each pair of nodes in the community will have a neighbor in common,
thus the radius of the subgraph is 2 and every node is a center
(with $r=2$). Thus the algorithm set with parameters $r=2$ is
(potentially) able to recover in phase I a superset of any
$\gamma$-quasi clique for $\gamma \geq 0.5$.

Consider the following random process. We take a set $V$ of $n$
vertices , for each $v \in V$ we select its neighbor vertex set by
picking vertices in $V$ with probability $p$. The property that any
two nodes have a non-empty neighbor intersection can be rephrased as
the requirement that the corresponding {\em intersection graph} (see
\cite{Karonski1999}) is complete. The threshold value of $p$
 so that this property holds with high probability is

 \[ p = \sqrt{\frac{2\log n}{n}} \]

Since the expected number of neighbors of any node $v$ is $pn$, we
have that the graph is an $O(p)$-quasi clique withe high
probability. Thus we are able to retain the property $r=2$ and the
fact  that all nodes are centers ever at a quite low overall
density.

\section{Testing effectiveness}
\label{sec:testing}
\label{sec:testing}

\subsection{Testing methodology}

The quality of the heuristic proposed will be tested empirically by
the planted solution methodology \cite{Feige05}. For a given input
graph $G$ and parameters $\delta$, and $q$  we implant randomly a
certain number of  subgraphs in $G$, by adding edges, so to force
the presence of a subgraph of size $\bar{q} \geq q$ and density
$\bar{\delta} \geq \delta$. We call this graph $G'$ the {\em planted
graph}.

By measuring  the number of such planted subgraphs that are found by
running $H-PDG$ blindly on the planted graph we can measure the
effectiveness of $H-PDG$  in finding such structures. In IR terms we
can thus measure the recall of the method. Experiments on 12 data
sets (not shown in this paper) indicate that using this simple
methodology the performance is close to 100\% in terms of recall for
planted cliques of size equal to the average degree of the graph
$G$. This methodology was adopted in \cite{DourisboureGP09}.

In order to better assess the potential of the method however we
resort to a more taxing testing methodology. In particular we want
to measure not only the algorithm's capability of detecting planted
dense subgraph, but also the property that no dense subgraphs are
left in the residual graph after the application of the algorithm.

Thus we operate as follows. We take the input graph $G$ and we apply
the $H-PDG$ algorithm, thus obtaining a set of dense subgraphs and a
residual graph $G_1$. We then apply a second time $H-PDG$ to $G_1$
possibly obtaining a few more communities and a second residual
graph $G_2$. Next we plant dense subgraphs in $G_2$ obtaining the
planted graph $G'_2$, and we run $H-PDG$ on $G'_2$. In the ideal
case $G_2$, being a residual graph after two applications of
$H-PDG$ is already empty of dense subgraphs, thus the all and only
communities we should find are those we planed in $G'_2$. Now we can
define properly the notion of {\em precision} and {\em recall} of
the algorithm. Each planted community that is not found reduces the
recall, while each found community that was not planted reduces the
precision. We can obviously also define the harmonic mean of the two
measures that is known as F-measure which gives us a more synthetic
measurement.

Note that we apply $H-PDG$ twice before generating the planting
graph on which we measure precision and recall. In general one can
try to increase the precision by iterating several time $H-PDG$ on
the residual graph resulting from the  previous iteration. We chose
to limit the iterations to a fixed and small number (two) for two
orders of reasons, the first one is that the cost of the execution
in terms of time  increases\footnote{This marginal cost could be
kept under control by doing incremental updates of the data
structures, rather than a re-computation from scratch and by
examining only the nodes whose immediate neighborhood has been
affected by the updates.}. The second and more important one is that
even with two applications the precision already reaches values
close to 1, while further iteration have no direct impact on the
recall, thus the F-measure would not be much affected by further
iterations.

\noindent {\bf Choice of parameters}.

The choice of size and density of the planted communities is
critical. If we plant  communities too large and dense the degree of
the selected nodes will be biased and even a simple minded approach
can easily detect the nodes belonging to these communities. We
choose the planted communities so that their average degree is equal
to the half the average degree of the host graph (i.e roughly
$|E|/|V|$).
 Moreover we choose the number of planted communities so to modify
no more than 1\% of the total number of nodes.

The searching algorithm is given as parameters (minimum size,
density) exactly the parameters used for the planting of the
communities. The list of detected communities is matched with the
list of embedded communities, and a planted community is marked as
"detected" if its vertex set overlaps by more than 50\% with a
detected community. Precision and recall are computed with respect
to the  communities (not their node-sets).

Note that one could improve the recall "by default" by performing
the search with parameters of minimum size and minimum density
slightly lower than those used for the generation. In this case we
are less likely to miss embedded nodes, and the embedded community
would be a subset of a larger detected community without incurring
in any penalty. In order to make a fair assessment of the proposed
algorithm we will not use this option. In general, however, such
strategy is beneficial when one sees $H-PDG$ as a filter to be used
to feed candidate communities (of much smaller size w.r.t the
original graph G) to exact but computationally expensive methods,
such as for example methods based on ILP developed in
\cite{Bala2007}.

\noindent {Other testing methodologies}. In \cite{Chen10} the
evaluation of the accuracy is done, for the case of non bipartite
graphs, by  generating  random graphs with 100 nodes and 2000 edges
that contains three dense subgraphs which cover 75\% of the nodes of
the graph. Precision and recall are computed with respect to the
node-sets and the   F-values obtained are close to 1.

In \cite{Wang2008} the evaluation of the accuracy is done by
embedding four dense subgraphs (with 8 nodes each and density 70\%)
in a random sparser graph of total of 60 nodes. Also, in this case
the  planted communities cover more than 50\% of the nodes of the
graph. In \cite{Wang2008} it is shown that all four dense subgraphs
are detected using the CSV-plot.

Although the results shown in \cite{Chen10,Wang2008} are quite good
for small random graphs with high community coverage, it is not
clear if measurements obtained allows to infer similar performance
in social graphs whose size, degree distribution and community
structure are so different.

For this reasons we decided to develop a new, more demanding,
testing methodology, which is closer to the applicative scenarios
where the targets are social and technological graphs.

\subsection{Benchmark data set}

The experiments are conducted on 25 graphs downloaded from the
Stanford Large Network Dataset Collection \cite{Stanford}. This
collection comprises several graph types including: Internet
autonomous systems, Authors collaboration networks, Internet
peer-to-peer networks, E-mail communication networks, Citation
networks, Web graphs, Trust networks, Road networks, and Product
co-purchasing networks. Basic graph statistics are shown in Table
\ref{ref:Table1}, other measures are available at \cite{Stanford}.

\subsection{Experiments}

In Tables \ref{ref:Table0} and \ref{ref:Table00} we report average
precision, average recall and average F-measure over all 25 data
sets, for density ranging from 100\% to 50\% and radius ranging from
1 to 2.

Tables \ref{ref:Table2}, \ref{ref:Table3}, \ref{ref:Table4}, and
\ref{ref:Table5} report  the precision, recall and fmeasure of the
method on each test file for varying radius and density values.

In the first set of experiment we embed cliques (100\% dense, radius
1) of size roughly $|E|/|V|$, data are reported in Table
\ref{ref:Table2}. Note that the recall is always quite high; it is
above 90\% in all 25 cases. Precision is slightly worse (above 90\%
in 22 cases out of 25). The combination though the F-measure however
is above 90\% in 23 cases out of 25, and never below 0.85. On
average, precision is about 96\% and recall about 99\%. When we keep
the radius to 1, but we reduce the density to 70\% (see
\ref{ref:Table3}), there is a reduction in precision and recall,
with recall above 90\% in 11 tests, precision above 90\% in 12
tests, with average precision  about 83\%  and average recall about
84\%. When we relax the radius constraint to radius=2 at 70\%
density, 14 test cases attain precision above 90\%, 10 recall above
90\%, and 8 F-measure above 90\%, however while the average recall
reduces to 78\%, the average precision is maintained at 85\%.
Finally when we leave the radius to 2 and set the density to 50\%
we have very similar values for average precision and average
recall. Thus, we conclude that the computation at radius 2 is not so
sensitive to the density threshold.

Tables \ref{ref:Table6}, \ref{ref:Table7}, \ref{ref:Table8}, and
\ref{ref:Table9} report on the time measurement on each test file,
with data for each phase, for the total time spent, and  the time
spent per edge. The initial phase that consists in the core
decomposition is linear in the size of the graph, while all other
computations involve the local neighborhood of some node, thus it is
interesting to measure the time spent per edge as a common criterion
of comparison for different runs of the algorithm. One can observe
that the time spend for edge is rather stable, in the region of
$10^-6$ to $10^-5$ seconds for radius =1
 (Tables \ref{ref:Table6}, \ref{ref:Table7}) and in the range
$10^-4$ to $10^-3$ seconds for radius=2 (Tables \ref{ref:Table8},
\ref{ref:Table9}). Some slower runs for radius=1 are noticeable for
some of the web-graphs (files 23, and 25). In the total time we
include the cost of run 3 whose purpose is to measure the
effectiveness of runs 1 and 2, thus, if this reliability measure is
not needed, it can be skipped and the total time reduced by about
1/3. Since the experimental code has been developed in Python we
expect that a factor 10 in timing can be easily gained by switching
to an implementation in  the C language.

\begin{table}[!h] \begin{center} \begin{tabular}{||r | l | l || r |
r | l ||} \hline measure & radius & density & value \\ \hline
average precision  & d=1 &  1.0 & 0.95429535884 \\ average recall
& d=1 & 1.0 & 0.98965535788\\ average fmeasure   & d=1 & 1.0 &
0.97020508204\\ \hline average precision  & d=1 & 0.7 &
0.83020137232\\ average recall     & d=1 & 0.7 & 0.83947183728\\
average fmeasure   & d=1 & 0.7 & 0.82681618136\\ \hline
\end{tabular} \end{center} \caption{Averages output quality
statistics for radius 1.} \label{ref:Table0} \end{table}

\begin{table}[!h] \begin{center} \begin{tabular}{||r | l | l || r |
r | l ||} \hline measure & radius & density & value \\ \hline
average precision  & d=2 &  0.7 &  0.8557872834 \\ average recall
& d=2 & 0.7 & 0.780538866 \\ average fmeasure   & d=2 & 0.7 &
0.78194706944\\ \hline average precision  & d=2 & 0.5 &
0.77783144912\\ average recall     & d=2 & 0.5 & 0.82669275188\\
average fmeasure   & d=2 & 0.5 & 0.78456921128\\ \hline
\end{tabular} \end{center} \caption{Averages output quality
statistics for radius 2.} \label{ref:Table00} \end{table}


\comment{ \begin{table*}[!h] \begin{center} \begin{tabular}{||r | l
| l || r | r | r ||} \hline \# & Type  & Graph Name & \# nodes & \#
edges & $2|E|/|V|$ \\ \hline \hline 1 & Autonomous System &
as20000102 & 6,474 & 26,467 &         8.17   \\ 2 & Autonomous
System  & oregon1\_010526 & 11,174  &  23,409 & 4.19  \\ 3 &
Autonomous System & oregon2\_010526 & 11,461 & 32,730  &   5.71\\ 4
& Autonomous System & as-caida20071112 & 26,475 & 106,762 &  8.06 \\
5 & Autonomous System  & as-skitter &  1,696,415 &  11,095,298 &
13.08   \\ \hline 6 & Collaboration Networks  & ca-GrQc  & 5,242  &
28,980    &  11.05  \\ 7 & Collaboration Networks  & ca-HepTh &
9,877  &  51,971   &   10.52     \\ 8 & Collaboration Networks  &
ca-HepPh & 12,008 &  237,010  &    39.47    \\ 9 & Collaboration
Networks  & ca-AstroPh & 18,772 &  396,160 &    42,20  \\ 10 &
Collaboration Networks  & ca-CondMat & 23,133 & 186,936   &   16.16
\\ \hline 11 & Citation Networks & cit-HepTh & 27,770  & 352,807 &
25.41  \\ 12 & Citation Networks & cit-HepPh & 34,546 & 841,798 &
49.47   \\ 13 & Citation Networks & cit-Patents & 3,774,768 &
16,518,948  &  8.75\\ \hline 14 & Web Graphs  & web-Stanford  &
281,903     &  2,312,497   &  16.39 \\ 15 & Web Graphs &
web-BerkStan  & 685,230     &  7,600,595   &  22.18 \\ 16 & Web
Graphs &  web-Google    & 875,713     &  5,105,039    &  11.67 \\
\hline 17 & Road Networks &  roadNet-PA.seq & 1,088,092  &
3,083,796 &  5.68 \\ 18 & Road Networks  & roadNet-TX.seq &
1,379,917 & 3,843,320 &  5.57   \\ 19 & Road Networks  &
roadNet-CA.seq & 1,965,206  & 5,533,214 & 5.63 \\ \hline 20 & Social
Network & wiki-Vote  & 7,115 & 201,524 &  56.64  \\ 21 & Social
Network & soc-Epinions1.seq  & 75,879 & 811,480 &  21.38   \\ 22 &
Social Network & soc-sign-Slashdot090221 &  82,144  & 549,202 &
13.37\\ 23  & Social Network & soc-Slashdot0922   & 82,168  &
948,464 &  23.08\\ \hline 24 & Communication Network & Email-Enron &
36,692 & 367,662 &  20.04   \\ 25 & Communication Network &
wiki-Talk &  2,394,385  &  5,021,410 &  4.19 \\ \hline 26 & Product
Co-Purchasing & amazon0601  & 403,394  &   3,387,388    &  16.79 \\
\hline 27 & Internet PeerToPeer & p2p-Gnutella31 & 62,586  & 147,892
&  4.72 \\ \hline \hline \end{tabular} \end{center} \caption{List of
27 graphs test from Stanford Large Network Dataset Collection with
number of nodes, number of edges and ratio $|E|/|V|$.}
\label{ref:Table1} \end{table*} }

\begin{table*}[!h] \begin{center} \begin{tabular}{||r | l | l || r |
r | r ||} \hline \# & Type  & Graph Name & \# nodes & \# edges &
$|E|/|V|$ \\ \hline \hline 1.  & AutonomousSystemGraphs    &
as-caida20071112      & 26389      &  105722   &  4,0062905 \\ 2.  &
AutonomousSystemGraphs    & as-skitter            & 1696415    &
22190596  &  13,08087703 \\ 3.  & AutonomousSystemGraphs    &
as20000102            & 6474       &25144      & 3,883843065 \\ 4.
& AutonomousSystemGraphs    & oregon1-010526   & 11174  &  46818  &
4,189905137 \\ 5.  & AutonomousSystemGraphs    & oregon2-010526   &
11461   & 65460  &  5,711543495\\ 6.  & CitationNetworks          &
cit-HepPh   &34546  &  841754   &24,36617843\\ 7.  &
CitationNetworks          & cit-HepTh   &27770   & 704570
&25,37162405\\ 8.  & CitationNetworks          & cit-Patents    &
3774768  &33037894  &   8,75229789\\
9. & CollaborationNetworks     & ca-CondMat     &23133   & 186878 &
8,078416116\\ 10. & CollaborationNetworks     & ca-GrQc   & 5242  &
28968   & 5,526135063\\ 11. & CollaborationNetworks     & ca-HepPh
&12008   & 236978  & 19,73500999\\ 12. & CollaborationNetworks     &
ca-HepTh   &9877   &  51946  &  5,259289258\\ 13. &
CommunicationNetwork      & Email-Enron    & 36692  &  367662 &
10,02022239\\ 14. & InternetPeerToPeer        & p2p-Gnutella31
&62586  &  295784  & 4,726040968\\ 15. & ProductCo-Purchasing      &
amazon0601  &403394   &4886816  &12,11425058\\ 16. & RoadNetworks
& roadNet-CA  &1965206 & 5533214  &2,815589816\\ 17. & RoadNetworks
& roadNet-PA  &1088092  &3083796  &2,834131673\\ 18. & RoadNetworks
& roadNet-TX  &1379917  &3843320  &2,785182007\\ 19. &
SignedNetworksP1          & soc-sign-Slashdot090221   &  82140   &
1000962 & 12,18604821\\ 20. & SignedNetworksP1          & wiki-Talk
& 2394385 & 9319130  &3,892076671\\
21. & SocialNetwork             & soc-Epinions1  &75879  &  811480
& 10,69439502\\ 22. & SocialNetwork             & soc-Slashdot0902
& 82168   & 1008460 & 12,2731477\\ 23. & WebGraphs                 &
web-BerkStan  & 685230   &13298940   &  19,4079944\\ 24. & WebGraphs
& web-Google    & 875713  & 8644102 & 9,870930316\\ 25. & WebGraphs
& web-Stanford   &281903   &3985272  &14,13703295\\ \hline \hline
\end{tabular} \end{center} \caption{List of 25 graphs test from
Stanford Large Network Dataset Collection with number of nodes,
number of edges and ratio $|E|/|V|$.} \label{ref:Table1}
\end{table*}


\comment{ \begin{table*}[!h] \begin{center} \begin{tabular}{|| r | l
| r | l || l | l | l ||} \hline \# & Graph Name &  size &  density
& precision & recall  & F-measure \\ \hline \hline 1 & as20000102
&  5  & 1.0  &   0.923076923  & 1.0         &   0.96 \\ 2 &
oregon1\_010526  &  5  & 1.0  &   0.733333333  & 1.0         &
0.846153846  \\ 3 & oregon2\_010526  &  6  & 1.0  &   0.655172414  &
1.0         &   0.791666667 \\ 4 & as-caida20071112 &  5  & 1.0  &
0.866666667  & 1.0         &   0.928571429 \\ 5 & as-skitter       &
14 (??)  & 1.0  &   0.996702391  & 0.998348472 &   0.997524752 \\
\hline 6 & ca-GrQc          & 6   & 1.0   &  0.888888889  & 1.0    &
0.941176471  \\ 7 & ca-HepTh         & 6   & 1.0   &  0.8          &
1.0    &   0.888888889 \\ 8 & ca-HepPh         & 20  & 1.0   & 1.0
& 1.0    &   1.0  \\
9 &ca-CondMat.      &  9  & 1.0   &  0.558139535  &  0.96   &
0.705882353 \\ \hline 10 & cit-HepTh       &  26 (?)  & 1.0   & 1.0
& 1.0      & 1.0 \\ 11 & cit-HepPh       &  25      & 1.0   & 1.0
& 1.0      & 1.0 \\ 12 & cit-Patents     & 9        & 1.0   & 1.0
& 1.0      & 1.0 \\ \hline 13  & web-Stanford    & 15 (?)   &  1.0 &
0.92       &  0.983957219  &   0.950904393 \\ 14  & web-BerkStan
& 20 (?)   & 1.0   &  0.988405797 & 0.997076023   &  0.99272198 \\
15  & web-Google      & 10  (?)  & 1.0   & 0.822456814  &
0.979428571    & 0.894105373 \\ \hline 16 & roadNet-PA     & 3  (?)
& 1.0  &    0.955119215  & 0.939051296 &   0,947017105 \\ 17 &
roadNet-TX     & 3   & 1.0    &    0.965747331  & 0.944118287  &
0.954810335 \\ 18 & roadNet-CA     & 3   & 1.0  &  0.950054799 &
0.926412214    &  0.938084564  \\ \hline
19 & soc-Epinions  &  11 & 1.0   & 1.0          & 1.0      & 1.0 \\
20 & soc-sign-Slashdot090221 &  13 & 1.0 &     0.984375  &  1.0    &
0.992125984 \\ 21 & soc-Slashdot0902 &   14 & 1.0   & 1.0          &
1.0      & 1.0 \\ \hline 22 & Email-Enron & 11 & 1.0 &
0.891891892 &  1.0  &        0.942857143  \\ 23 & wiki-Talk   &   4
&  1.0  &    0.911832236  & 0.995321637  & 0.951749481  \\ \hline 24
& amazon0601  &  13  (?) & 1.0   & 1.0          & 1.0      & 1.0 \\
\hline 25 & p2p-Gnutella31   &  5 & 1.0   & 1.0          & 1.0
& 1.0 \\ \hline \hline \end{tabular} \end{center}
\caption{Experiments with embedding of communities 100\% dense of
size about $|E|/|V|$. We report the size of the communities
embedded, the recall precision and f-measure of our algorithm.}
\label{ref:Table2} \end{table*} }


\begin{table*}[!h] \begin{center} \begin{tabular}{|| r | l | r |l |
l || l | l | l | r ||} \hline \# & Graph Name &  \# nodes &  density
& av. degree  & precision & recall  & F-measure &  Num. Embedded\\
\hline \hline 1. & as-caida20071112 &  5 &  1  & 4  & 0,928571429  &
0,99047619 &  0,958525346  & 105 \\ 2. & as-skitter  & 14 & 1  & 13
&0,998351195 &0,999587288 &0,99896886  &2423\\ 3. & as20000102  & 5
& 1 &  4  & 0,961538462 &1   &0,980392157& 25\\ 4. & oregon1-010526
&  5  & 1  & 4  & 0,843137255& 0,977272727 &0,905263158 &44\\ 5. &
oregon2-010526  & 6  & 1  & 5  & 0,791666667 &1  & 0,88372093  &38\\
6. & cit-HepPh & 25 & 1  & 24 & 1  & 1  & 1  & 27\\ 7. & cit-HepTh &
26 & 1  & 25 & 1  & 1 &  1  & 21\\ 8. & cit-Patents   & 9  & 1   &8
& 1 &  1 &  1 &  8388\\
9 & ca-CondMat  &  9   &1  & 8 &  0,739130435& 1  & 0,85  &  51\\ 10
& ca-GrQc  & 6 &  1  & 5 &  1  & 1  & 1  & 17\\ 11  & ca-HepPh & 20
& 1 &  19 & 1 &  1 &  1 &  12\\ 12 & ca-HepTh & 6  & 1  & 5  &
0,914285714 &1  & 0,955223881 &32\\ 13 & Email-Enron  &  11 & 1  &
10 & 0,942857143 &1  & 0,970588235 &66\\ 14 & p2p-Gnutella31   &5  &
1  & 4  & 1  & 1 &  1  & 250\\ 15 & amazon0601 & 13 & 1  & 12 & 1
&1 &  1  & 620\\ 16 & roadNet-CA & 4  & 1 &  3  & 0,974968112&
0,933516525& 0,953792162 &13101\\ 17 & roadNet-PA  & 4  & 1  & 3 &
0,976884422 &0,938094582 &0,957096638 &7253\\ 18 & roadNet-TX  & 4
& 1  & 3  & 0,982360923 &0,944450484 &0,963032755 &9199\\ 19 &
soc-sign-Slashdot090221  &  13  &1   &12  &0,992063492 &0,992063492&
0,992063492& 126\\ 20 & wiki-Talk & 4   &1  & 34  & 0,954465429
&0,996324451& 0,974945845 &11971\\
21 & soc-Epinions1 &11  &1   &10  &1  & 1  & 1  & 137\\ 22 &
soc-Slashdot0902  &14  &1  & 13 & 1  & 1  & 1  & 117\\ 23 &
web-BerkStan  &20 & 1  & 19  &0,994186047 &0,998540146 &0,996358339
&685\\ 24 & web-Google   & 10  &1   &9 & 0,904260915 &0,981724729
&0,941401972& 1751\\ 25 & web-Stanford  &15  &1  & 14  &0,958656331
&0,989333333 &0,973753281 &375\\ \hline \hline \end{tabular}
\end{center} \caption{Experiments with embedding of communities of
radius 1, 100\% dense of average degree about $|E|/|V|$. We report
the size of the communities embedded, the recall precision and
f-measure of our algorithm.} \label{ref:Table2} \end{table*}

\begin{table*}[!h] \begin{center} \begin{tabular}{|| r | l | r |l |
l || l | l | l | r ||} \hline \# & Graph Name &  \# nodes  &
density & av. degree & precision & recall  & F-measure &  Num.
Embedded\\ \hline \hline 1. & as-caida20071112 & 7 &  0,7 & 5  &
0,905405405 &0,893333333 &0,899328859 &75\\

2. & as-skitter  & 20 & 0,7 &14 & 0,940600522 &0,849646226
&0,892812887 &1696\\

3 . & as20000102  & 7  & 0,7 &5  & 0,857142857 &1   &0,923076923
&18\\

4. & oregon1-010526 &  7  & 0,7 & 5  & 0,783783784 &0,935483871
&0,852941176 &31\\

5. & oregon2-010526   &8  & 0,7 &6  & 0,631578947 &0,857142857
&0,727272727& 28\\

6. & cit-HepPh  &35 & 0,7 &25 & 1  & 0,947368421 &0,972972973 &19\\

7. & cit-HepTh & 37  &0,7 &26 & 0,8 &0,533333333 &0,64  &  15\\

8. & cit-Patents    &12  &0,7& 9  & 0,972490988 &0,814814815
&0,886697803 &6291\\


9 & ca-CondMat   & 12 & 0,7 &9  & 0,444444444 &0,736842105
&0,554455446 &38\\

10. & ca-GrQc  & 8   &0,7& 6  & 0,866666667 &1  & 0,928571429 &13\\

11. & ca-HepPh  &28  &0,7 &20 & 0,3 &0,375  & 0,333333333 &8\\

12. & ca-HepTh & 8  & 0,7 &6  & 0,75  &  1   &0,857142857& 24\\

13. & Email-Enron   & 15 & 0,7 &11 & 0,612903226 &0,791666667
&0,690909091 &48\\

14. & p2p-Gnutella31  & 7  & 0,7 &5  & 1  & 0,786516854 &0,880503145
&178\\

15. & amazon0601 &18  &0,7 &13  &0,965517241 &1   &0,98245614
&448\\

16. & roadNet-CA &4   &0,7 &3   &1  & 0,984225524& 0,992050059
&9826\\

17. & roadNet-PA &4   &0,7 & 3  & 1  & 0,984375   & 0,992125984
&5440\\

18. & roadNet-TX &4  & 0,7 &3  & 0,999705623& 0,984490506
&0,992039728 &6899\\

19. & soc-sign-Slashdot090221   & 18  &0,7 &13  &0,984848485&
0,714285714& 0,828025478 &91\\

20. & wiki-Talk  &5   &0,7 &4  & 0,923212253 &0,931502558
&0,927338877 &9577\\


21 & soc-Epinions1 &15  &0,7 &11  &0,775510204 &0,752475248
&0,763819095 &101\\

22 & soc-Slashdot0902  &20  &0,7 &14  &0,971014493& 0,817073171
&0,887417219 &82\\

23 & web-BerkStan  &28  &0,7 &20  &0,872641509 &0,756646217
&0,810514786 &489\\

24 & web-Google    &14  &0,7 &10 & 0,576139089 &0,768185452
&0,658444673& 1251\\

25 & web-Stanford & 21 & 0,7 &15  &0,821428571 &0,77238806 &
0,796153846 & 268\\ \hline \hline \end{tabular} \end{center}
\caption{Experiments with embedding of communities of radius 1, 70\%
dense of average degree about $|E|/|V|$. We report the size of the
communities embedded, the recall precision and f-measure of our
algorithm.} \label{ref:Table3} \end{table*}


\begin{table*}[!h] \begin{center} \begin{tabular}{|| r | l | r |l |
l || l | l | l | r ||} \hline \# & Graph Name &  \# nodes &  density
& av. degree  & precision & recall  & F-measure &  Num. Embedded\\
\hline \hline 1. & as-caida20071112 & 7  & 0,7& 5 &  0,948717949
&0,986666667 &0,967320261 &75\\

2. & as-skitter &  20 & 0,7 &14 & 0,918563923 &0,618514151&
0,739252995 &1696\\

3. & as20000102  & 7  & 0,7& 5 &  0,947368421& 1 &  0,972972973&
18\\

4. & oregon1-010526  & 7   &0,7& 5  & 0,885714286 &1 &  0,939393939
&31\\

5. & oregon2-010526  & 8 &  0,7& 6  & 0,675  & 0,964285714&
0,794117647& 28\\

6. & cit-HepPh & 35  &0,7& 25 & 1  & 0,684210526 &0,8125 & 19\\

7. & cit-HepTh  &37 & 0,7 &26 & 1  & 0,2& 0,333333333 &15\\

8. & cit-Patents  &  12 & 0,7 &9  & 0,979404193 &0,846606263&
0,908176315 &6291\\


9. & ca-CondMat   & 12 & 0,7 &9   &0,421875  &  0,710526316&
0,529411765 &38\\

10. & ca-GrQc  & 8  & 0,7 &6 &  0,666666667 &0,923076923&
0,774193548 &13\\

11. & ca-HepPh  &28  &0,7 &20 & 0,583333333 &0,875 &  0,7 &8\\

12. & ca-HepTh & 8  & 0,7  & 6  & 0,8 &1  & 0,888888889& 24\\

13. & Email-Enron   & 15 & 0,7& 11  &0,634920635 &0,833333333
&0,720720721 &48\\

14. & p2p-Gnutella31  & 7 &  0,7& 5  & 1  & 0,578651685 &0,733096085
&178\\

15. & amazon0601 & 18  &0,7 &13 & 0,991150442 &1  & 0,995555556
&448\\

16. & roadNet-CA & 4  & 0,7& 3  & 0,99922179  &0,914716059
&0,955103342& 9826\\

17. & roadNet-PA & 4  & 0,7& 3 &  0,999193873 &0,911397059
&0,953278216 &5440\\

18. & roadNet-TX  & 4  & 0,7 &3  & 0,99904943 & 0,914045514
&0,954658996 &6899\\

19. & soc-sign-Slashdot090221  &  18 & 0,7 &13 & 0,925925926
&0,274725275& 0,423728814 &91\\

20. & wiki-Talk & 5  & 0,7 &4  & 0,922119447 &0,843165918
&0,880877059 &9577\\


21. & soc-Epinions1& 15 & 0,7 &11 & 0,788235294& 0,663366337
&0,720430108 &101\\

22. & soc-Slashdot0902  &20 & 0,7 &14 & 0,923076923& 0,292682927
&0,444444444 &82\\

23. & web-BerkStan  &28 & 0,7 &20 & 0,884520885& 0,736196319
&0,803571429 &489\\

24 & web-Google    &14  &0,7& 10 & 0,64084507 & 0,872901679
&0,739086294 &1251\\

25. & web-Stanford & 21 & 0,7 &15 & 0,859778598 &0,869402985
&0,864564007 &268\\

\hline \hline \end{tabular} \end{center} \caption{Experiments with
embedding of communities of radius 2, 70\% dense of average degree
about $|E|/|V|$. We report the size of the communities embedded, the
recall precision and f-measure of our algorithm.} \label{ref:Table4}
\end{table*}


\begin{table*}[!h] \begin{center} \begin{tabular}{|| r | l | r |l |
l || l | l | l | r ||} \hline \# & Graph Name &  \# nodes &  density
& av. degree  & precision & recall  & F-measure &  Num. Embedded\\
\hline \hline

1. & as-caida20071112 & 10  & 0,5  &5  &  0,894736842  &0,980769231
&0,935779817  &52\\

2. & as-skitter   & 28   &0,5  &14   &0,770140429  &0,860445912
&0,812792512  &1211\\

3. & as20000102   & 10  & 0,5  &5   & 0,8 & 1   & 0,888888889  &12\\

4. & oregon1-010526   & 10  & 0,5  &5  &  0,846153846  &1   &
0,916666667  &22\\

5. & oregon2-010526    &12  & 0,5 & 6   & 0,464285714  &0,684210526
&0,553191489  &19\\

6. & cit-HepPh  & 50   &0,5  &25  & 1   & 1    &1   & 13\\

7. & cit-HepTh   &52  & 0,5  &26  & 0,714285714  &0,5 & 0,588235294
&10\\

8. & cit-Patents    & 18 &  0,5  &9   & 0,873723104  &0,95851216  &
0,91415577  & 4194\\


9. & ca-CondMat    & 18  & 0,5 & 9   & 0,46   &  0,92  &
0,613333333  &25\\

10. & ca-GrQc    &12  & 0,5  &6   & 0,727272727  &1   & 0,842105263
&8\\

11. & ca-HepPh  & 40  & 0,5 & 20  & 0,25   &  0,333333333
&0,285714286  &6\\

12. & ca-HepTh   &12  & 0,5  &6    &0,842105263 & 1   & 0,914285714
&16\\

13. & Email-Enron    & 22  & 0,5  &11  &0,5  &0,787878788
&0,611764706  &33\\

14. & p2p-Gnutella31    &10  & 0,5  &5   & 1    &0,696  &
0,820754717  &125\\

15. & amazon0601  &26   &0,5  &13  & 0,987261146  &1   & 0,993589744
&310\\

16. & roadNet-CA  &6   & 0,5  &3   & 0,99908383  & 0,998931298 &
0,999007558  &6550\\

17. & roadNet-PA  &6   & 0,5 & 3   & 0,99917287  & 0,999448428
&0,99931063   &3626\\

18. & roadNet-TX  &6   & 0,5  &3   & 0,999130435  &0,999347684
&0,999239048  &4599\\

19. & soc-sign-Slashdot090221    & 26  & 0,5  &13  & 0,942857143
&0,523809524  &0,673469388  &63\\

20. & wiki-Talk   &8  &  0,5 & 4   & 0,941370768  &0,952380952
&0,946843854  &5985\\


21. & soc-Epinions1  &22  & 0,5  &11  & 0,776315789  &0,867647059
&0,819444444  &68\\

22. & soc-Slashdot0902  & 28  & 0,5 & 14  & 0,962962963
&0,448275862  &0,611764706  &58\\

23. & web-BerkStan   &40  & 0,5  &20  & 0,62915601  & 0,719298246
&0,671214188  &342\\

24. & web-Google     &20  & 0,5 & 10  & 0,374229346  &0,693714286
&0,48618342   &875\\

25. & web-Stanford   &30  & 0,5  &15  & 0,691542289 & 0,743315508
&0,716494845  &187\\

\hline \hline \end{tabular} \end{center} \caption{Experiments with
embedding of communities of radius 2, 50\% dense of average degree
about $|E|/|V|$. We report the size of the communities embedded, the
recall precision and f-measure of our algorithm.} \label{ref:Table5}
\end{table*}



\begin{table*}[!h] \begin{center} \begin{tabular}{|| r | l | r || r
|r | r || r | r ||} \hline \hline num.  & file name & numarcs & time
ph1 & time ph2 & time ph3 & total time & cost per arc \\ \hline 1  &
as-caida20071112  &  105722  &  0.34  &  0.32  &  0.34  &  1  &
9.46e-06  \\ 2  &  as-skitter  &  22190596  &  181.12  &  373.33  &
414.16  &  968.61  &  4.36e-05  \\ 3  &  as20000102  &  25144  &
0.06  &  0.08  &  0.1  &  0.24  &  9.55e-06  \\ 4  &  oregon1-010526
&  46818  &  0.11  &  0.12  &  0.13  &  0.36  &  7.69e-06  \\ 5  &
oregon2-010526  &  65460  &  0.14  &  0.21  &  0.23  &  0.58  &
8.86e-06  \\ 6  &  cit-HepPh  &  841754  &  3.3  &  3.42  &  3.31  &
10.03  &  1.19e-05  \\ 7  &  cit-HepTh  &  704570  &  3.75  &  3.94
&  3.88  &  11.57  &  1.64e-05  \\ 8  &  cit-Patents  &  33037894  &
721.13  &  786  &  797.23  &  2304.36  &  6.97e-05  \\ 9  &
ca-CondMat  &  186878  &  0.46  &  0.54  &  0.58  &  1.58  &
8.45e-06  \\ 10  &  ca-GrQc  &  28968  &  0.07  &  0.07  &  0.07  &
0.21  &  7.25e-06  \\ 11  &  ca-HepPh  &  236978  &  0.3  &  0.46  &
0.6  &  1.36  &  5.74e-06  \\ 12  &  ca-HepTh  &  51946  &  0.17  &
0.17  &  0.18  &  0.52  &  1.00e-05  \\ 13  &  Email-Enron  &
367662  &  1.1  &  1.57  &  1.83  &  4.5  &  1.22e-05  \\ 14  &
p2p-Gnutella31  &  295784  &  1.12  &  1.22  &  1.25  &  3.59  &
1.21e-05  \\ 15  &  amazon0601  &  4886816  &  28.04  &  28.27  &
27.37  &  83.68  &  1.71e-05  \\ 16  &  roadNet-CA  &  5533214  &
107.29  &  99.73  &  100.2  &  307.22  &  5.55e-05  \\ 17  &
roadNet-PA  &  3083796  &  45.23  &  41.92  &  43.51  &  130.66  &
4.24e-05  \\ 18  &  roadNet-TX  &  3843320  &  63.58  &  59.38  &
58.61  &  181.57  &  4.72e-05  \\ 19  &  soc-sign-Slashdot090221  &
1000962  &  3.95  &  4.37  &  4.57  &  12.89  &  1.29e-05  \\ 20  &
wiki-Talk  &  9319130  &  198.96  &  213.97  &  165.91  &  578.84  &
6.21e-05  \\ 21  &  soc-Epinions1  &  811480  &  3.43  &  4.14  &
4.67  &  12.24  &  1.51e-05  \\ 22  &  soc-Slashdot0902  &  1008460
&  3.9  &  4.7  &  4.79  &  13.39  &  1.33e-05  \\ 23  &
web-BerkStan  &  13298940  &  1066.51  &  2897.22  &  2890.32  &
6854.05  &  5.15e-04  \\ 24  &  web-Google  &  8644102  &  61.88  &
75.49  &  82.16  &  219.53  &  2.54e-05  \\ 25  &  web-Stanford  &
3985272  &  91.5  &  106.75  &  268.76  &  467.01  &  1.17e-04  \\
\hline \hline \end{tabular} \end{center} \caption{Experiments with
embedding of communities of radius 1, 100\% density. Timing of
phase,1, 2 and 3 and total time (in seconds), and time pr arc.}
\label{ref:Table6} \end{table*}


\begin{table*}[!h] \begin{center} \begin{tabular}{|| r | l | r || r
|r | r || r | r ||} \hline \hline num.  & file name & numarcs & time
ph1 & time ph2 & time ph3 & total time & cost per arc \\ \hline 1  &
as-caida20071112  &  105722  &  0.3  &  0.3  &  0.33  &  0.93  &
8.80e-06  \\ 2  &  as-skitter  &  22190596  &  159.25  &  192.38  &
203.75  &  555.38  &  2.50e-05  \\ 3  &  as20000102  &  25144  &
0.06  &  0.06  &  0.06  &  0.18  &  7.16e-06  \\ 4  &
oregon1-010526  &  46818  &  0.11  &  0.11  &  0.12  &  0.34  &
7.26e-06  \\ 5  &  oregon2-010526  &  65460  &  0.14  &  0.15  &
0.16  &  0.45  &  6.87e-06  \\ 6  &  cit-HepPh  &  841754  &  2.08
&  2.42  &  2.45  &  6.95  &  8.26e-06  \\ 7  &  cit-HepTh  &
704570  &  1.84  &  2.07  &  2.53  &  6.44  &  9.14e-06  \\ 8  &
cit-Patents  &  33037894  &  702.65  &  765.72  &  786.36  &
2254.73  &  6.82e-05  \\ 9  &  ca-CondMat  &  186878  &  0.4  &  0.4
&  0.44  &  1.24  &  6.64e-06  \\ 10  &  ca-GrQc  &  28968  &  0.06
&  0.06  &  0.06  &  0.18  &  6.21e-06  \\ 11  &  ca-HepPh  &
236978  &  0.29  &  0.25  &  0.27  &  0.81  &  3.42e-06  \\ 12  &
ca-HepTh  &  51946  &  0.14  &  0.14  &  0.15  &  0.43  &  8.28e-06
\\ 13  &  Email-Enron  &  367662  &  0.84  &  0.92  &  0.98  &  2.74
&  7.45e-06  \\ 14  &  p2p-Gnutella31  &  295784  &  1.08  &  1.17
&  1.22  &  3.47  &  1.17e-05  \\ 15  &  amazon0601  &  4886816  &
23.04  &  23.04  &  22.56  &  68.64  &  1.40e-05  \\ 16  &
roadNet-CA  &  5533214  &  103.42  &  100.03  &  99.82  &  303.27  &
5.48e-05  \\ 17  &  roadNet-PA  &  3083796  &  45.93  &  43.61  &
44.2  &  133.74  &  4.34e-05  \\ 18  &  roadNet-TX  &  3843320  &
60.25  &  57.78  &  58.66  &  176.69  &  4.60e-05  \\ 19  &
oc-sign-Slashdot090221  &  1000962  &  3.4  &  4.11  &  4.23  &
11.74  &  1.17e-05  \\ 20  &  wiki-Talk  &  9319130  &  128.42  &
142.71  &  154.92  &  426.05  &  4.57e-05  \\ 21  &  soc-Epinions1
&  811480  &  2.31  &  2.81  &  3.37  &  8.49  &  1.05e-05  \\ 22  &
soc-Slashdot0902  &  1008460  &  3.32  &  3.88  &  3.96  &  11.16  &
1.11e-05  \\ 23  &  web-BerkStan  &  13298940  &  65.43  &  103.16
&  99.81  &  268.4  &  2.02e-05  \\ 24  &  web-Google  &  8644102  &
56.76  &  53.59  &  54.02  &  164.37  &  1.90e-05  \\ 25  &
web-Stanford  &  3985272  &  43.93  &  47.11  &  60.47  &  151.51  &
3.80e-05  \\ \hline \hline \end{tabular} \end{center}
\caption{Experiments with embedding of communities of radius 1, 70\%
density. Timing of phase,1, 2 and 3, total time (in seconds), and
time per arc.} \label{ref:Table7} \end{table*}



\begin{table*}[!h] \begin{center} \begin{tabular}{|| r | l | r || r
|r | r || r | r ||} \hline \hline num.  & file name & numarcs & time
ph1 & time ph2 & time ph3 & total time & cost per arc \\ \hline 1  &
as-caida20071112  &  105722  &  2.94  &  18.64  &  20.17  &  41.75
&  3.95e-04  \\ 2  &  as-skitter  &  22190596  &  10902.6  &
19621.01  &  25241.08  &  55764.69  &  2.51e-03  \\ 3  &  as20000102
&  25144  &  0.51  &  1.83  &  1.49  &  3.83  &  1.52e-04  \\ 4  &
oregon1-010526  &  46818  &  0.96  &  3.62  &  1.87  &  6.45  &
1.38e-04  \\ 5  &  oregon2-010526  &  65460  &  1.08  &  3.24  &
2.78  &  7.1  &  1.08e-04  \\ 6  &  cit-HepPh  &  841754  &  68.42
&  88.93  &  89.38  &  246.73  &  2.93e-04  \\ 7  &  cit-HepTh  &
704570  &  68.98  &  161.12  &  170.39  &  400.49  &  5.68e-04  \\ 8
&  cit-Patents  &  33037894  &  1796  &  2077.75  &  2150.96  &
6024.71  &  1.82e-04  \\ 9  &  ca-CondMat  &  186878  &  5.53  &
6.72  &  7.02  &  19.27  &  1.03e-04  \\ 10  &  ca-GrQc  &  28968  &
0.48  &  0.75  &  0.76  &  1.99  &  6.87e-05  \\ 11  &  ca-HepPh  &
236978  &  3.26  &  9.68  &  11.01  &  23.95  &  1.01e-04  \\ 12  &
ca-HepTh  &  51946  &  1.44  &  1.78  &  1.75  &  4.97  &  9.57e-05
\\ 13  &  Email-Enron  &  367662  &  13.42  &  23.38  &  40.11  &
76.91  &  2.09e-04  \\ 14  &  p2p-Gnutella31  &  295784  &  9.48  &
9.6  &  9.62  &  28.7  &  9.70e-05  \\ 15  &  amazon0601  &  4886816
&  322.33  &  317.33  &  310.45  &  950.11  &  1.94e-04  \\ 16  &
roadNet-CA  &  5533214  &  237.53  &  234.08  &  227.45  &  699.06
&  1.26e-04  \\ 17  &  roadNet-PA  &  3083796  &  120.03  &  117.52
&  115.2  &  352.75  &  1.14e-04  \\ 18  &  roadNet-TX  &  3843320
&  152.14  &  149.46  &  145.01  &  446.61  &  1.16e-04  \\ 19  &
soc-sign-Slashdot090221  &  1000962  &  359.46  &  418.74  &  428.14
&  1206.34  &  1.21e-03  \\ 20  &  wiki-Talk  &  9319130  &
14539.25  &  15883.42  &  6385.09  &  36807.76  &  3.95e-03  \\ 21
&  soc-Epinions1  &  811480  &  169.48  &  233.11  &  250.23  &
652.82  &  8.04e-04  \\ 22  &  soc-Slashdot0902  &  1008460  &
375.55  &  424.08  &  429.07  &  1228.7  &  1.22e-03  \\ 23  &
web-BerkStan  &  13298940  &  10322.09  &  21415.46  &  15577.4  &
47314.95  &  3.56e-03  \\ 24  &  web-Google  &  8644102  &  241.02
&  445.79  &  510.39  &  1197.2  &  1.38e-04  \\ 25  &  web-Stanford
&  3985272  &  2122.91  &  3970.01  &  3123.08  &  9216  &  2.31e-03
\\ \hline \hline \end{tabular} \end{center} \caption{Experiments
with embedding of communities of radius 2, 70\%. Timing of phase,1,
2 and 3 and total time (in seconds), and time per arc.}
\label{ref:Table8} \end{table*}


\begin{table*}[!h] \begin{center} \begin{tabular}{|| r | l | r || r
|r | r || r | r ||} \hline \hline num.  & file name & numarcs & time
ph1 & time ph2 & time ph3 & total time & cost per arc \\ \hline 1  &
as-caida20071112  &  105722  &  2.11  &  7.06  &  5.77  &  14.94  &
1.41e-04  \\ 2  &  as-skitter  &  22190596  &  2212  &  4198.62  &
5365.61  &  11776.23  &  5.31e-04  \\ 3  &  as20000102  &  25144  &
0.44  &  0.57  &  0.6  &  1.61  &  6.40e-05  \\ 4  &  oregon1-010526
&  46818  &  0.74  &  1.28  &  1.29  &  3.31  &  7.07e-05  \\ 5  &
oregon2-010526  &  65460  &  0.82  &  1.25  &  1.31  &  3.38  &
5.16e-05  \\ 6  &  cit-HepPh  &  841754  &  44.63  &  73.97  &
72.63  &  191.23  &  2.27e-04  \\ 7  &  cit-HepTh  &  704570  &
35.52  &  83.87  &  124.16  &  243.55  &  3.46e-04  \\ 8  &
cit-Patents  &  33037894  &  1426.51  &  1564.77  &  1584.58  &
4575.86  &  1.39e-04  \\ 9  &  ca-CondMat  &  186878  &  4.57  &
5.43  &  5.78  &  15.78  &  8.44e-05  \\ 10  &  ca-GrQc  &  28968  &
0.39  &  0.56  &  0.59  &  1.54  &  5.32e-05  \\ 11  &  ca-HepPh  &
236978  &  1.85  &  3.78  &  4.06  &  9.69  &  4.09e-05  \\ 12  &
ca-HepTh  &  51946  &  1.13  &  1.4  &  1.42  &  3.95  &  7.60e-05
\\ 13  &  Email-Enron  &  367662  &  8.48  &  13.71  &  29.75  &
51.94  &  1.41e-04  \\ 14  &  p2p-Gnutella31  &  295784  &  8.2  &
8.6  &  8.63  &  25.43  &  8.60e-05  \\ 15  &  amazon0601  &
4886816  &  263.66  &  283.89  &  272.97  &  820.52  &  1.68e-04  \\
16  &  roadNet-CA  &  5533214  &  200.26  &  196.21  &  191.29  &
587.76  &  1.06e-04  \\ 17  &  roadNet-PA  &  3083796  &  99.3  &
97.69  &  93.52  &  290.51  &  9.42e-05  \\ 18  &  roadNet-TX  &
3843320  &  125.83  &  124.1  &  120.3  &  370.23  &  9.63e-05  \\
19  &  oc-sign-Slashdot090221  &  1000962  &  339.9  &  404.43  &
408.17  &  1152.5  &  1.15e-03  \\ 20  &  wiki-Talk  &  9319130  &
14793.02  &  15704.76  &  6636.37  &  37134.15  &  3.98e-03  \\ 21
&  soc-Epinions1  &  811480  &  93.41  &  158.88  &  261.85  &
514.14  &  6.34e-04  \\ 22  &  soc-Slashdot0902  &  1008460  &
341.68  &  404.75  &  412.42  &  1158.85  &  1.15e-03  \\ 23  &
web-BerkStan  &  13298940  &  5382.16  &  5311.22  &  5286.71  &
15980.09  &  1.20e-03  \\ 24  &  web-Google  &  8644102  &  177.88
&  223.74  &  250.02  &  651.64  &  7.54e-05  \\ 25  &  web-Stanford
&  3985272  &  1133.52  &  1859.58  &  1889.28  &  4882.38  &
1.23e-03  \\ \hline \hline \end{tabular} \end{center}
\caption{Experiments with embedding of communities of radius 2,
50\%. Timing of phase,1, 2 and 3 and total time (in seconds) and
time per arc.} \label{ref:Table9} \end{table*}

\section{Conclusions}
\label{sec_conclusions}

In this paper we have presented a method to quickly find dense
sub-graphs (communities) in graphs. The novelty of the approach
relies on the combination of existing methods to speed up the
community discovery. The experiments conducted on various types of
graphs show the effectiveness and the reliability of the method.

\balance

\bibliographystyle{abbrv}

\end{document}